# $Cr_2O_3/\beta\text{-}Ga_2O_3$ Heterojunction Diodes with Orientation-Dependent Breakdown Electric Field up to 12.9 MV/cm

Yizheng Liu[1,a)], Haochen Wang[1], Carl Peterson[1], James S. Speck[1], Chris Van De Walle[1], and Sriram Krishnamoorthy[1,a)]

[1]Materials Department, University of California Santa Barbara, Santa Barbara CA 93106, USA

a) Author(s) to whom correspondence should be addressed. Electronic mail: *yizhengliu@ucsb.edu*,, *sriramkrishnamoorthy@ucsb.edu*

**Abstract**: We report the fabrication of $Cr_2O_3/\beta\text{-}Ga_2O_3$ heterojunction diodes using reactive magnetron sputtering of $Cr_2O_3$ on highly doped $\beta\text{-}Ga_2O_3$ bulk substrates along (100), (010), (001), (110), and (011) orientation dependence of high electric field handling capability in $\beta\text{-}Ga_2O_3$. Additional relative permittivity values in (110) and (011) orientations of $\beta\text{-}Ga_2O_3$ were computed by using first-principles calculation methods for accurate apparent charge density ($N_D-N_A$) extraction and breakdown electric field analysis from capacitance-voltage measurements. The HJDs fabricated on $n^+$ (110) exhibited breakdown electric fields >10 MV/cm up to 12.9 MV/cm, showing the highest experimentally observed parallel-plane junction electric field among $\beta\text{-}Ga_2O_3$-based junctions. Breakdown electric fields among (100), (010), (001), and (011) orientations showed distinct distribution in the range of 5.13-5.26 MV/cm, 5.10-7.05 MV/cm, 2.70-3.33 MV/cm, and 3.88-4.38 MV/cm, respectively, validating the orientational dependence of parallel-plane junction electric field at breakdown in low-symmetry monoclinic $\beta\text{-}Ga_2O_3$. The parallel-plane breakdown electric fields ($E_{Br,\parallel}$) reported in this work were extracted when the device experienced catastrophic breakdown at 100 mA/cm$^2$ current density compliance, and should not be confused with critical electric field ($E_c$) as a function of drift layer doping concentration, which accounts for electric-field dependent impact ionization coefficients in Si, SiC and GaN. This study can guide the choice of crystal orientation for high performance gallium oxide-based devices that require high electric field handling capability.

In recent years of wide/ultra-wide bandgap (WBG/UWBG) semiconductor advancements, $\beta\text{-}Ga_2O_3$ offers promising potential for medium-voltage power applications (1-35 kV) in applications of power conversion at grid scale, renewable energy processing, and data centers for artificial intelligence (AI). The high critical electric field strength and availability of the shallow hydrogenic dopants[1] in epitaxial $\beta\text{-}Ga_2O_3$[2–5] can be leveraged to demonstrate power devices with much lower differential specific on-resistance while operating at a higher blocking voltage compared to silicon carbide (SiC) and gallium nitride (GaN)[6,7]. To fully exploit the high-electric field advantage of $\beta\text{-}Ga_2O_3$, integrations with high-dielectric constant (high-κ) oxides[8,9] were realized to achieve 5.7 MV/cm average electric field at breakdown[10]. High electric field of 4.3 MV/cm has been reported using oxidized large barrier height $\beta\text{-}Ga_2O_3$ Schottky barrier diodes (SBDs)[11]. Although reliable p-type doping with mobile holes in $\beta\text{-}Ga_2O_3$ is currently unavailable, p-type oxides, such as nickel oxide (NiO), was integrated with $\beta\text{-}Ga_2O_3$ to form P-N heterojunctions that readily enable multi-kV junctions[12–17] with breakdown voltage >10 kV[18–20] and high electric field at 7.5 MV/cm[21]. Because of the monoclinic asymmetric crystal structure of β-$Ga_2O_3$[22], anisotropies in various material parameters are expected, such as relative permittvity[23,24], thermal conductivity[25], and critical electric field[26]. Recently, we have reported crystal orientation-dependence of breakdown electric field between (100) and (001) β-$Ga_2O_3$ using NiO/β-$Ga_2O_3$ diodes[21]. Nevertheless, a systematic comparison of breakdown electric field analysis across available crystal orientations of bulk-grown β-$Ga_2O_3$ needs to be explored.

Similar to NiO, chromium oxide ($Cr_2O_3$), an intrinsic p-type material[27] with a bandgap of ~3.6 eV[28,29], has recently been explored to form P-N heterojunctions[30–32] with β-$Ga_2O_3$ as an alternative to NiO-based junctions, exhibiting enhanced thermal stability at high temperatures (600 °C)[32].

In this work, we examine $Cr_2O_3$/β-$Ga_2O_3$ heterojunction diodes (HJDs) via reactive magnetron sputtering on several orientations ((100), (010), (001), (110), and (110) $n^+$ β-$Ga_2O_3$ bulk crystals) to further examine the breakdown electric field ($E_{Br,\parallel}$) anisotropy and explore potential orientations that withstand high electric field. Extraction of breakdown fields require accurate characterization of doping profiles and dielectric constants along each of these crystal orientations. In this work, we report relative permittivity values for (110) and (011) crystal orientations which we can obtain from first-principles calculations. All five orientations of β-$Ga_2O_3$ on which HJDs are fabricated exhibit distinct breakdown electric field anisotropy with (110)-plane orientation showing $E_{Br,\parallel}$ > 10 MV/cm with the highest experimentally observed parallel-plane junction electric field at 12.9 MV/cm in β-$Ga_2O_3$, further validating the orientational breakdown electric field dependence in low-symmetry monoclinic β-$Ga_2O_3$. One important aspect to mention is that the parallel-plane breakdown electric fields ($E_{Br,\parallel}$) reported in this work were extracted when the device experienced catastrophic breakdown at 100 mA/$cm^2$ current density compliance, and should not be confused with critical electric field ($E_c$) as a function of drift layer doping concentration, which accounts for electric-field dependent impact ionization coefficients in Si, SiC and GaN[33].

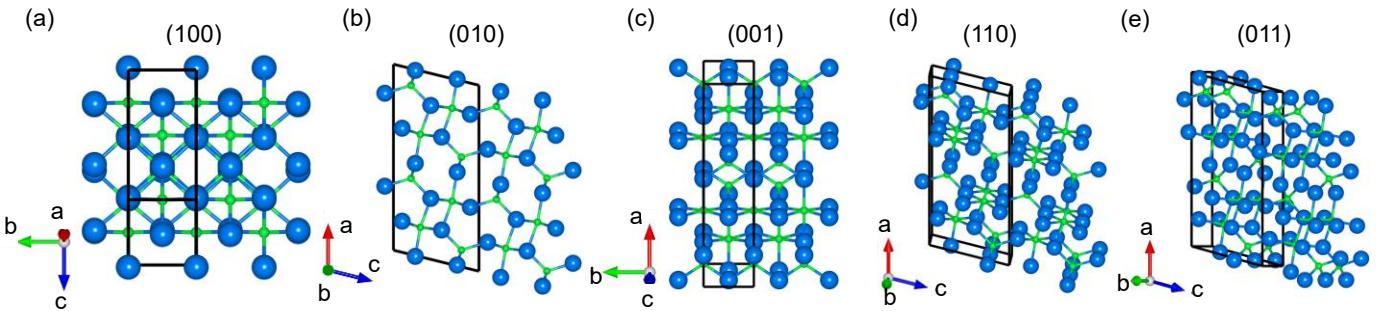

FIG.1. (a)-(e) Single layer VESTA schematics of monoclinic β-$Ga_2O_3$ crystals in (100), (010), (001), (110), (011) orientations with solid lines enclosed unit cells, respectively.

The monoclinic β-Ga₂O₃ crystal exhibits low symmetry, and this is visualized in **Fig. 1(a)-1(e)** by using Visualization for Electronic and Structural Analysis (VESTA) generated crystal structures projected along (100), (010), (001), (110), (011) orientations (one monolayer), indicating distinct atomic arrangement and highly varying in-plane atomic density.

To extract the apparent doping density and analyze breakdown electric field systematically in each orientation of β-Ga₂O₃ substrates, knowledge of the orientation-dependent relative permittivity is essential. The list of relative permittivity values in various orientations of β-Ga₂O₃ is shown in Table. I, obtained from first-principles calculations.

Table I. Dielectric permittivity along directions perpendicular to different lattice planes of β-Ga₂O₃, from both first-principles calculations and experiments.

| Lattice plane | $\varepsilon_{\text{eff}}$ (DFT) | $\varepsilon_{\text{eff}}$ (Experiment[23]) |
|---|---|---|
| (100) | 10.25 | 10.2±0.2 |
| (010) | 10.09 | 10.87±0.08 |
| (001) | 13.19 | 12.4±0.4 |
| (110) | 10.10 | - |
| (011) | 10.79 | - |
| (101) | 12.17 | - |

We calculated permittivity values based on density functional perturbation theory (DFPT)[34], using projector augmented wave potentials[35,36] as implemented in the Vienna *ab initio* simulation package (VASP)[37]. The generalized gradient approximation (GGA) by Perdew, Burke, and Ernzerhof (PBE)[38] is employed. We used a 20-atom conventional supercell with a $6 \times 10 \times 4$ Gamma-centered k-point set and a plane-wave cutoff energy of 500 eV. The structure was fully optimized until forces converged within 5 meV/Å. The resulting lattice vectors, in cartesian coordinates, are

$$\mathbf{a} = \begin{pmatrix} a_0 \\ 0 \\ 0 \end{pmatrix}, \mathbf{b} = \begin{pmatrix} 0 \\ b_0 \\ 0 \end{pmatrix}, \mathbf{c} = \begin{pmatrix} c_0 \\ 0 \\ c_1 \end{pmatrix} \quad (1)$$

where $a_0 = 12.51$ Å, $b_0 = 3.10$ Å, $c_0 = -1.40$ Å and $c_1 = 5.75$ Å, corresponding to lattice parameters $|\mathbf{a}| = 12.51$ Å, $|\mathbf{b}| = 3.10$ Å and $|\mathbf{c}| = 5.92$ Å. The angle between **c** and the z axis is 13.74°.

The dielectric permittivity tensor in this coordinate system is

$$\varepsilon_{ij}^{\text{tot}} = \begin{pmatrix} 10.52 & 0 & -0.90 \\ 0 & 10.09 & 0 \\ -0.90 & 0 & 13.19 \end{pmatrix}. \quad (2)$$

When an external electric field is applied on a monoclinic crystal, the induced electric displacement **D** is collinear only if the external field is along one of three principal axes. For instance, an electric field along **a** will induce electric displacements along x and z axes simultaneously. In the case of β-Ga₂O₃, only the lattice vector **b** is along one principal axis.

To determine the dielectric permittivity $\varepsilon'$ along an arbitrary direction, the electric displacement vector must be projected onto the direction of external field $\hat{\mathbf{n}} = (n_1, n_2, n_3)^T$:

$$\mathbf{D} \cdot \hat{\mathbf{n}} = \varepsilon' |\mathbf{E}| \tag{3}$$

. Combined with $D_i = \varepsilon_{ij} E_j$, we derive the dielectric permittivity along $\hat{\mathbf{n}}$:

$$\varepsilon' = \varepsilon_{11} n_1^2 + \varepsilon_{22} n_2^2 + \varepsilon_{33} n_3^2 + 2\varepsilon_{13} n_1 n_3 \tag{4}$$

where $\varepsilon_{11}, \varepsilon_{22}, \varepsilon_{33}$ and $\varepsilon_{13}$ are independent components of the dielectric tensor.

To compare with experimental values[21], we calculated $\varepsilon'$ along directions normal to specific lattice planes. For the (100) plane, $\hat{\mathbf{n}} = \frac{1}{K}(b_0, a_0, -c_0 b_0/c_1)$ with $K = \sqrt{b_0^2 + a_0^2 + c_0^2 b_0^2/c_1^2}$, resulting in $\varepsilon' = 10.25$, in good agreement with the experimental value $10.2 \pm 0.2$ (Table I). The results for the (010), (001), (110), and (011) orientations are also included in Table I.

Schottky barrier diodes (SBDs), as shown in **Fig. 2(a)**, were fabricated on highly doped β-Ga₂O₃ substates along five orientations to extract the accurate apparent charge density. The SBDs were fabricated using electron-beam (e-beam) evaporated Ni/Au (50/150 nm) Schottky contacts and annealed e-beam evaporated Ti/Au (50/200 nm) backside Ohmic contacts. Ohmic contact annealing was performed at 470 °C in nitrogen (N₂) ambient for 60 seconds. Capacitance-voltage (C-V) measurements were performed using an AC frequency of 1 MHz (Fig. S1(a) in the supplementary materials) to extract the apparent charge density in the five substrates along different orientations.

The Cr₂O₃/ β-Ga₂O₃ heterojunction diode (HJD) fabrication began with a backside Ti/Au (50/150 nm) Ohmic metallization using e-beam evaporation followed by a 60-seconds rapid thermal annealing (RTA) at 470 °C in N₂. Following the backside Ohmic contact formation, a photoresist lift-off mask was patterned using optical lithography after standard solvent clean (acetone/isopropanol/de-ionized water). Then, a p⁻ Cr₂O₃ layer ($R_{sheet}$ >1 MΩ/□) was directly deposited on all five orientations of n⁺ β-Ga₂O₃ via reactive radio frequency (RF) magnetron sputtering by using a 99.95%-pure metallic chromium (Cr) target under an oxygen-deficient condition (Ar/O₂~45/5 sccm) at a 3.7 mTorr chamber pressure and 150 W RF power. To enhance the Ohmic contact quality to Cr₂O₃, a p⁺ Cr₂O₃ contact layer ($R_{sheet}$ ~600 kΩ/□) was sputtered under an oxygen-rich condition (Ar/O₂~8/10 sccm) at a 3 mTorr chamber pressure and the same RF power. Immediately following the sputter deposition, a Ni/Au/Ni (50/50/200 nm) metal stack was deposited via e-beam evaporation to serve as an Ohmic contact to Cr₂O₃ and a hard mask for subsequent plasma dry etching[21,39]. The Cr₂O₃/Ohmic metal stack was later lifted off in a heated N-methyl pyrrolidone (NMP) solution. To mitigate the electric field crowding, the diodes were dry-etched ~1.2 μm below the Cr₂O₃/β-Ga₂O₃ heterojunction interface under inductively coupled

BCl$_3$ plasma at 200 W, as shown in **Fig. 2(b)**. This self-aligned plasma etch served as edge termination for the device.

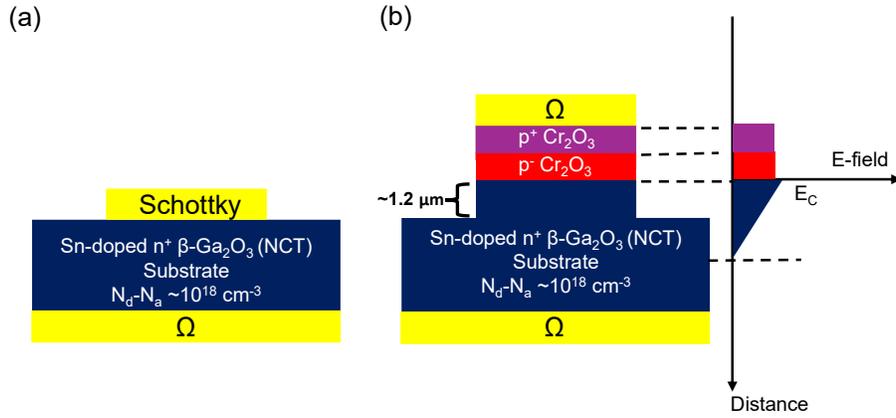

FIG.2. (a) n$^+$ β-Ga$_2$O$_3$ Schottky barrier diode schematic. (b) Cr$_2$O$_3$/β-Ga$_2$O$_3$ heterojunction diode on highly doped bulk substrate with electric field profile described on the side.

Capacitance voltage characterization was performed on SBDs, and the built-in potential was extracted from the 1/C$^2$ vs. Voltage characteristics at 1 MHz, as shown in **Fig. 3(a)**. The built-in potential extracted on the five orientations range from 1.12 to 1.43 V. The net apparent charge densities of highly doped β-Ga$_2$O$_3$ in all five orientations were extracted from C-V measurements to be 3.20×10$^{18}$ cm$^{-3}$ (100), 1.89×10$^{18}$ cm$^{-3}$ (010), 3.96×10$^{18}$ cm$^{-3}$ (001), 6.65×10$^{18}$ cm$^{-3}$ (110), and 3.56×10$^{18}$ cm$^{-3}$ (011), respectively, as shown in **Fig. 3(b)**, using DFPT-calculated relative permittivity values (Table. I) for various β-Ga$_2$O$_3$ orientations.

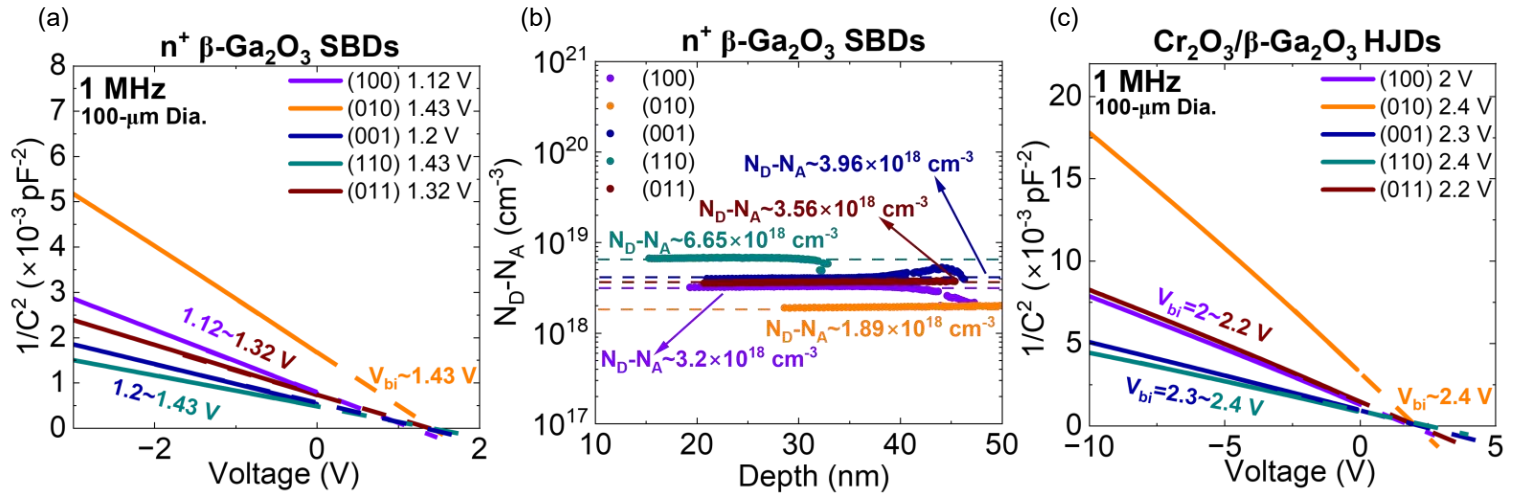

FIG.3. (a) 1/C$^2$ vs. Voltage characteristics of fabricated n$^+$ β-Ga$_2$O$_3$ Schottky barrier diode with extracted built-in potential values. (b) Apparent charge density vs. depth profile of n$^+$ β-Ga$_2$O$_3$ bulk substrates in all five orientations extracted from n$^+$ β-Ga$_2$O$_3$ SBD C-V characteristics. (c) 1/C$^2$ vs. Voltage characteristics of fabricated Cr$_2$O$_3$/β-Ga$_2$O$_3$ heterojunction diode on highly doped bulk substrate with extracted built-in potential values.

The built-in potential values for the HJDs were extracted from $1/C^2$ vs. Voltage characteristics at 1 MHz as shown in **Fig. 3(c)**. The extracted built -in potential ranged from 2 to 2.4 V, confirming the formation of heterojunction between sputtered $Cr_2O_3$ and $n^+$ $\beta$-$Ga_2O_3$. The detailed capacitance vs. voltage characteristics of fabricated $Cr_2O_3$/$\beta$-$Ga_2O_3$ HJDs are shown in S1(b) along with their J-V characteristics (Fig.S2(a)-2(b)) and SBDs' J-V characteristics (Fig.S2(c)-2(d)) in both linear and semi-log scales in the supplementary materials.

The $Cr_2O_3$/$\beta$-$Ga_2O_3$ HJDs' reverse leakage current density and breakdown voltage characteristics are shown in **Fig. 4(a)-4(e)** for all five orientations of $\beta$-$Ga_2O_3$. By using a 100 mA/cm$^2$ current density as the compliance (JEDEC standards)[21] for non-catastrophic leakage-mediated device breakdown, the HJDs' reverse breakdown voltages were extracted to be 31.5-33 V, 47.5-86.0 V, 13-18.5 V, 95-108.5 V, and 25.5-31 V on (100), (010), (001), (110), and (011) $n^+$ $\beta$-$Ga_2O_3$, respectively. Since the $Cr_2O_3$/$\beta$-$Ga_2O_3$ HJDs were fabricated on highly doped substrates, the depletion width under reverse bias into the $n^+$ $\beta$-$Ga_2O_3$ can be comparable to the total thickness of the sputtered $Cr_2O_3$. Therefore, a triangular non-punch through electric field profile, as shown in **Fig. 2(b)**, was assumed. To extract the lower-bound value of the parallel-plane junction breakdown electric field, both sputtered $p^-$ and $p^+$ $Cr_2O_3$ layers were assumed to be fully depleted with rectangular electric field profiles. Therefore, the total voltage at device breakdown (100 mA/cm$^2$) can be expressed as

$$t_{Cr_2O_3}^{total} \times \frac{q(N_D-N_A)W_n}{\varepsilon_{s(Cr_2O_3)}} + \frac{1}{2} \times \frac{q(N_D-N_A)W_n^2}{\varepsilon_{s(\beta-Ga_2O_3)}} = V_{100\ mA/cm^2} + V_{bi} \qquad (5)$$

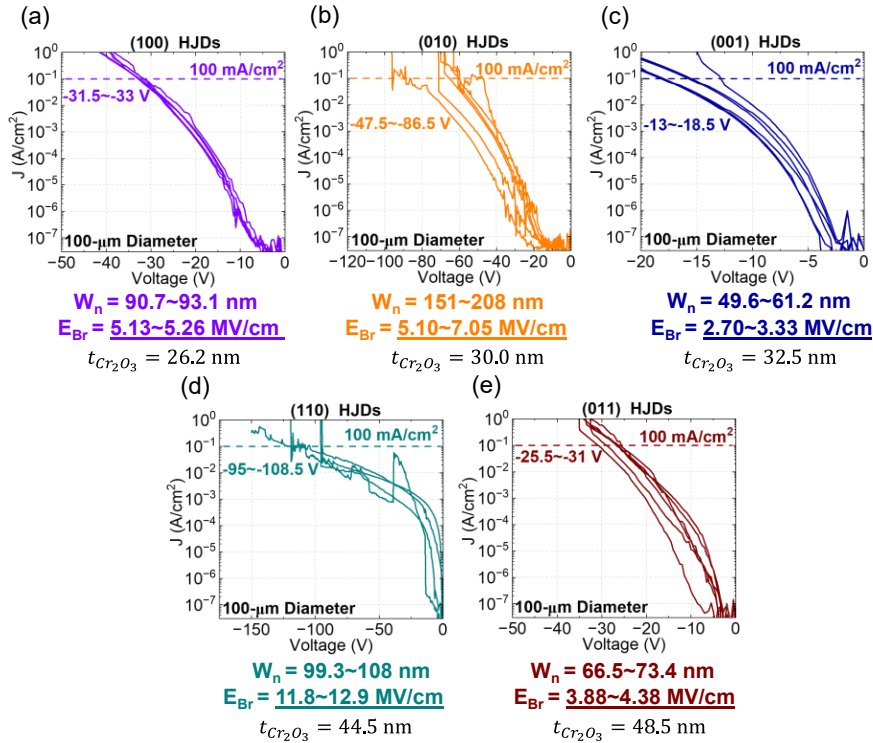
(a) (100) HJDs, 100-μm Diameter, -31.5~-33 V, $W_n$ = 90.7~93.1 nm, $E_{Br}$ = 5.13~5.26 MV/cm, $t_{Cr_2O_3}$ = 26.2 nm

(b) (010) HJDs, 100-μm Diameter, -47.5~-86.5 V, $W_n$ = 151~208 nm, $E_{Br}$ = 5.10~7.05 MV/cm, $t_{Cr_2O_3}$ = 30.0 nm

(c) (001) HJDs, 100-μm Diameter, -13~-18.5 V, $W_n$ = 49.6~61.2 nm, $E_{Br}$ = 2.70~3.33 MV/cm, $t_{Cr_2O_3}$ = 32.5 nm

(d) (110) HJDs, 100-μm Diameter, -95~-108.5 V, $W_n$ = 99.3~108 nm, $E_{Br}$ = 11.8~12.9 MV/cm, $t_{Cr_2O_3}$ = 44.5 nm

(e) (011) HJDs, 100-μm Diameter, -25.5~-31 V, $W_n$ = 66.5~73.4 nm, $E_{Br}$ = 3.88~4.38 MV/cm, $t_{Cr_2O_3}$ = 48.5 nm

FIG.4. Reverse breakdown characteristics of $Cr_2O_3/\beta$-$Ga_2O_3$ heterojunction diodes on (a) (100), (b) (010), (c) (001), (d) (110), and (e) (011) bulk $n^+$ $\beta$-$Ga_2O_3$ substrates with calculated corresponding depletion depth and parallel-plane junction electric field at device breakdown.

Assuming a relative permittivity of 13.42[40] for $Cr_2O_3$ ($\varepsilon_{s(Cr_2O_3)}$), and 10.25 for (100) $\beta$-$Ga_2O_3$ ($\varepsilon_{s(\beta-Ga_2O_3)}$) from the calculated DFPT results in Table. I with a total combined $Cr_2O_3$ thickness ($t_{Cr_2O_3}^{total}$) of 26.2 nm on $n^+$ (100) $\beta$-$Ga_2O_3$ and apparent charge density of $3.2 \times 10^{18}$ cm$^{-3}$ ($N_D$-$N_A$) with 2 V built-in potential, the depletion width ($W_n$) at reverse bias breakdown into the $n^+$ (100) $\beta$-$Ga_2O_3$ was calculated to be 90.7-93.1 nm, which corresponds to a parallel-plane junction electric field ($E_{Br,\parallel}$) of 5.13-5.26 MV/cm at breakdown using the following relation

$$E_{Br,\parallel} = \frac{q(N_D - N_A)W_n}{\varepsilon_{s(\beta-Ga_2O_3)}} \tag{6}$$

Similarly, the $W_n$ of HJDs at breakdown on (010), (001), (110), and (011) $n^+$ substrates were extracted to be 151-208 nm, 49.6-61.2 nm, 99.3-108 nm, and 66.5-73.4 nm with corresponding calculated $E_{Br,\parallel}$ at 5.10-7.05 MV/cm, 2.70-3.33 MV/cm, 11.8-12.9 MV/cm, and 3.88-4.38 MV/cm, respectively, as shown in **Fig. 4(b)-4(e)**. The extracted breakdown electric fields of $Cr_2O_3$-based HJDs on $n^+$ (100) and (001) closely match results on NiO-based HJDs[21], and further validate breakdown field anisotropy observed in this work. Distribution of the extracted $E_{Br,\parallel}$ for the five orientations investigated in this work was plotted in **Fig. 5(a)**, indicating a strong anisotropy in electric field handling capability for different orientations of $\beta$-$Ga_2O_3$. The origin of such a large anisotropy remains to be investigated. A maximum parallel-plane junction electric field ($E_{Br,\parallel}$) value of 12.9 MV/cm was extracted from HJDs fabricated on $n^+$ (110) $\beta$-$Ga_2O_3$. We report the highest experimentally observed $E_{Br,\parallel}$ in $\beta$-$Ga_2O_3$-based junctions, as shown in the benchmark[10–12,18,21,39,41] in **Fig. 5(b)**.

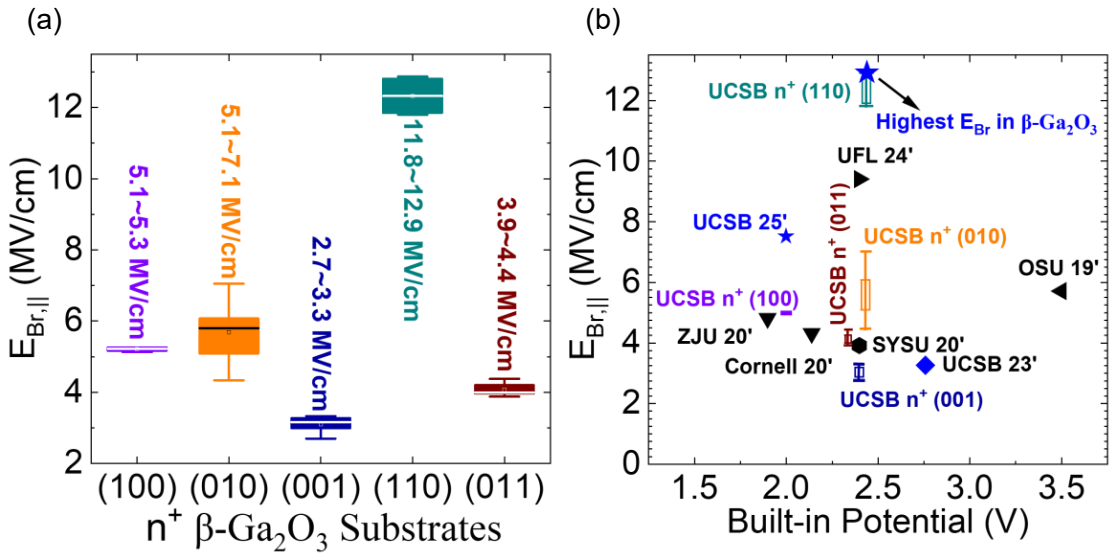

FIG.5. (a) $E_{Br,\parallel}$ distribution at breakdown for n$^+$ (100), (010), (001), (110), and (100) β-Ga$_2$O$_3$ substrates. (b) Benchmarking of $E_{Br,\parallel}$ for β-Ga$_2$O$_3$ vertical power devices[10–12,18,39,41].

We also quantified the uncertainty of the highest parallel-plane breakdown electric field reported on n$^+$ (110) β-Ga$_2$O$_3$ in this work. With a 30% uncertainty in the relative permittivity of Cr$_2$O$_3$ ($\varepsilon_{(Cr2O3)}$ = 9.4~17.4), the calculated lower bound $E_{Br,\parallel}$ was 11.5-13.5 MV/cm (4.7~10.9% variation compared to the original calculated results, $E_{Br,\parallel}$=12.9 MV/cm). The parallel-plane breakdown electric fields ($E_{Br,\parallel}$) reported in this work were extracted when the device experienced catastrophic breakdown at 100 mA/cm$^2$ current density compliance, and should not be confused with critical electric field ($E_c$) as a function of drift layer doping concentration, which accounts for electric-field dependent impact ionization coefficients in Si, SiC and GaN[33].

In summary, this work demonstrates the orientation-dependent breakdown electric field anisotropy in Cr$_2$O$_3$/β-Ga$_2$O$_3$ heterojunction diodes. Relative permittivity values along (110) and (011) orientations of β-Ga$_2$O$_3$ are obtained from first-principles calculations. The 12.9 MV/cm parallel-plane breakdown electric field obtained on (110) β-Ga$_2$O$_3$ exhibits the highest experimentally observed breakdown electric field among β-Ga$_2$O$_3$-based junctions across literatures.

**ACKNOWLEDGMENTS**

The authors acknowledge funding from the U.S. Department of Energy (DOE) ARPA-E ULTRAFAST program (DE-AR0001824) and Coherent/II-VI Foundation Block Gift Program. A portion of this work was performed at the UCSB Nanofabrication Facility, an open access laboratory.

**SUPPLEMENTARY MATERIAL**

See the supplementary material for a list of calculated relative permittivity values of β-Ga$_2$O$_3$ in desired orientations using first-principles methods, and the capacitance vs. voltage characteristics and J-V characteristics of both n$^+$ β-Ga$_2$O$_3$ Schottky barrier diode and Cr$_2$O$_3$/β-Ga$_2$O$_3$ heterojunction diode on highly doped bulk substrates.

**DATA AVAILABILITY**

The data that supports the findings of this study are available from the corresponding authors upon reasonable request.

# REFERENCES

[1] A.T. Neal, S. Mou, S. Rafique, H. Zhao, E. Ahmadi, J.S. Speck, K.T. Stevens, J.D. Blevins, D.B. Thomson, and N. Moser, "Donors and deep acceptors in β-Ga$_2$O$_3$," Applied Physics Letters **113**(6), (2018).

[2] A. Bhattacharyya, C. Peterson, K. Chanchaiworawit, S. Roy, Y. Liu, S. Rebollo, and S. Krishnamoorthy, "Over 6 μm thick MOCVD-grown low-background carrier density ($10^{15}$ cm$^{-3}$) high-mobility (010) β-Ga$_2$O$_3$ drift layers," Applied Physics Letters **124**(1), (2024).

[3] C. Peterson, A. Bhattacharyya, K. Chanchaiworawit, R. Kahler, S. Roy, Y. Liu, S. Rebollo, A. Kallistova, T.E. Mates, and S. Krishnamoorthy, "200 cm$^2$/Vs electron mobility and controlled low $10^{15}$ cm$^{-3}$ Si doping in (010) β-Ga$_2$O$_3$ epitaxial drift layers," Applied Physics Letters **125**(18), (2024).

[4] S. Rebollo, Y. Liu, C. Peterson, S. Krishnamoorthy, and J.S. Speck, "Growth of nitrogen-doped (010) β-Ga$_2$O$_3$ by plasma-assisted molecular beam epitaxy using an O$_2$/N$_2$ gas mixture," Applied Physics Letters **126**(8), (2025).

[5] C. Peterson, C.N. Saha, R. Kahler, Y. Liu, A. Mattapalli, S. Roy, and S. Krishnamoorthy, "Kilovolt-class β-Ga$_2$O$_3$ field-plated Schottky barrier diodes with MOCVD-grown intentionally $10^{15}$ cm$^{-3}$ doped drift layers," Journal of Applied Physics **138**(18), (2025).

[6] Y. Zhang, and J.S. Speck, "Importance of shallow hydrogenic dopants and material purity of ultra-wide bandgap semiconductors for vertical power electron devices," Semiconductor Science and Technology **35**(12), 125018 (2020).

[7] B.J. Baliga, *Fundamentals of Power Semiconductor Devices* (Springer Science & Business Media, 2010).

[8] S. Roy, B. Kostroun, J. Cooke, Y. Liu, A. Bhattacharyya, C. Peterson, B. Sensale-Rodriguez, and S. Krishnamoorthy, "Ultra-low reverse leakage in large area kilo-volt class β-Ga$_2$O$_3$ trench Schottky barrier diode with high-k dielectric RESURF," Applied Physics Letters **123**(24), (2023).

[9] S. Roy, B. Kostroun, Y. Liu, J. Cooke, A. Bhattacharyya, C. Peterson, B. Sensale-Rodriguez, and S. Krishnamoorthy, "Low Q$_C$ V$_F$ 20A/1.4 kV β-Ga$_2$O$_3$ Vertical Trench High-k RESURF Schottky Barrier Diode with Turn-on Voltage of 0.5 V," IEEE Electron Device Letters, (2024).

[10] Z. Xia, H. Chandrasekar, W. Moore, C. Wang, A.J. Lee, J. McGlone, N.K. Kalarickal, A. Arehart, S. Ringel, F. Yang, and S. Rajan, "Metal/BaTiO$_3$/β-Ga$_2$O$_3$ dielectric heterojunction diode with 5.7 MV/cm breakdown field," Applied Physics Letters **115**(25), 252104 (2019).

[11] D. Saraswat, W. Li, K. Nomoto, D. Jena, and H.G. Xing, "Very High Parallel-Plane Surface Electric Field of 4.3 MV/cm in Ga2O3 Schottky Barrier Diodes with PtOx Contacts," in *2020 Device Research Conference (DRC)*, (2020), pp. 1–2.

[12] J. Wan, H. Wang, C. Zhang, Y. Li, C. Wang, H. Cheng, J. Li, N. Ren, Q. Guo, and K. Sheng, "3.3 kV-class NiO/β-Ga$_2$O$_3$ heterojunction diode and its off-state leakage mechanism," Applied Physics Letters **124**(24), (2024).

[13] H.H. Gong, X.H. Chen, Y. Xu, F.-F. Ren, S.L. Gu, and J.D. Ye, "A 1.86-kV double-layered NiO/β-Ga$_2$O$_3$ vertical p–n heterojunction diode," Applied Physics Letters **117**(2), (2020).

[14] J.-S. Li, C.-C. Chiang, X. Xia, H.-H. Wan, F. Ren, and S.J. Pearton, "7.5 kV, 6.2 GW cm$^{-2}$ NiO/β-Ga$_2$O$_3$ vertical rectifiers with on–off ratio greater than $10^{13}$," Journal of Vacuum Science & Technology A **41**(3), (2023).

[15] J.-S. Li, H.-H. Wan, C.-C. Chiang, T.J. Yoo, F. Ren, H. Kim, and S.J. Pearton, "NiO/Ga$_2$O$_3$ Vertical Rectifiers of 7 kV and 1 mm$^2$ with 5.5 A Forward Conduction Current," Crystals **13**(12), 1624 (2023).

[16] M. Xiao, B. Wang, J. Spencer, Y. Qin, M. Porter, Y. Ma, Y. Wang, K. Sasaki, M. Tadjer, and Y. Zhang, "NiO junction termination extension for high-voltage (> 3 kV) Ga$_2$O$_3$ devices," Applied Physics Letters **122**(18), (2023).

[17] F. Zhou, H. Gong, M. Xiao, Y. Ma, Z. Wang, X. Yu, L. Li, L. Fu, H.H. Tan, and Y. Yang, "An avalanche-and-surge robust ultrawide-bandgap heterojunction for power electronics," Nature Communications **14**(1), 4459 (2023).

[18] J.-S. Li, H.-H. Wan, C.-C. Chiang, T.J. Yoo, M.-H. Yu, F. Ren, H. Kim, Y.-T. Liao, and S.J. Pearton, "Breakdown up to 13.5 kV in NiO/β-Ga$_2$O$_3$ vertical heterojunction rectifiers," ECS Journal of Solid State Science and Technology **13**(3), 035003 (2024).

[19] Y. Qin, Z. Yang, H. Gong, A.G. Jacobs, J. Spencer, M. Porter, B. Wang, K. Sasaki, C.-H. Lin, and M. Tadjer, "10 kV, 250° C Operational, Enhancement-Mode Ga$_2$O$_3$ JFET with Charge-Balance and Hybrid-Drain Designs," in *2024 IEEE International Electron Devices Meeting (IEDM)*, (IEEE, 2024), pp. 1–4.

[20] Y. Qin, M. Xiao, M. Porter, Y. Ma, J. Spencer, Z. Du, A.G. Jacobs, K. Sasaki, H. Wang, and M. Tadjer, "10-kV Ga$_2$O$_3$ charge-balance Schottky rectifier operational at 200° C," IEEE Electron Device Letters **44**(8), 1268–1271 (2023).

[21] Y. Liu, S.M. Witsell, J.F. Conley, and S. Krishnamoorthy, "Orientation-dependent β-Ga$_2$O$_3$ heterojunction diode with atomic layer deposition (ALD) NiO," Applied Physics Letters **127**(12), (2025).

[22] A.J. Green, J. Speck, G. Xing, P. Moens, F. Allerstam, K. Gumaelius, T. Neyer, A. Arias-Purdue, V. Mehrotra, and A. Kuramata, "β-Gallium oxide power electronics," APL Materials **10**(2), 029201 (2022).

[23] A. Fiedler, R. Schewski, Z. Galazka, and K. Irmscher, "Static dielectric constant of β-Ga$_2$O$_3$ perpendicular to the principal planes (100),(010), and (001)," ECS Journal of Solid State Science and Technology **8**(7), Q3083 (2019).

[24] P. Gopalan, S. Knight, A. Chanana, M. Stokey, P. Ranga, M.A. Scarpulla, S. Krishnamoorthy, V. Darakchieva, Z. Galazka, K. Irmscher, A. Fiedler, S. Blair, M. Schubert, and B. Sensale-Rodriguez, "The anisotropic quasi-static permittivity of single-crystal β-Ga$_2$O$_3$ measured by terahertz spectroscopy," Appl. Phys. Lett. **117**(25), 252103 (2020).

[25] Z. Guo, A. Verma, X. Wu, F. Sun, A. Hickman, T. Masui, A. Kuramata, M. Higashiwaki, D. Jena, and T. Luo, "Anisotropic thermal conductivity in single crystal β-gallium oxide," Applied Physics Letters **106**(11), (2015).

[26] K. Ghosh, and U. Singisetti, "Impact ionization in β-Ga$_2$O$_3$," Journal of Applied Physics **124**(8), (2018).

[27] M.T. Greiner, M.G. Helander, W.-M. Tang, Z.-B. Wang, J. Qiu, and Z.-H. Lu, "Universal energy-level alignment of molecules on metal oxides," Nature Materials **11**(1), 76–81 (2012).

[28] S. Ghosh, M. Baral, R. Kamparath, R.J. Choudhary, D.M. Phase, S.D. Singh, and T. Ganguli, "Epitaxial growth and interface band alignment studies of all oxide α-Cr$_2$O$_3$/β-Ga$_2$O$_3$ pn heterojunction," Applied Physics Letters **115**(6), (2019).

[29] H. Mashiko, T. Oshima, and A. Ohtomo, "Band-gap narrowing in α-(Cr$_x$Fe$_{1-x}$)$_2$O$_3$ solid-solution films," Applied Physics Letters **99**(24), (2011).

[30] C. Su, H. Zhou, Z. Hu, C. Wang, Y. Hao, and J. Zhang, "1.96 kV p-Cr$_2$O$_3$/β-Ga$_2$O$_3$ heterojunction diodes with an ideality factor of 1.07," Appl. Phys. Lett. **126**(13), 132104 (2025).

[31] Y. Feng, H. Zhou, J. Ma, H. Fang, X. Zhang, Y. Chen, G. Tian, J. Yuan, R. Peng, S. Xu, Y. Hao, and J. Zhang, "120 A/1.78 kV p-$Cr_2O_3$/n-β-$Ga_2O_3$ Heterojunction PN Diodes With Slanted Mesa Edge Termination," IEEE Electron Device Letters **46**(10), 1705–1708 (2025).

[32] W.A. Callahan, K. Egbo, C.-W. Lee, D. Ginley, R. O'Hayre, and A. Zakutayev, "Reliable operation of $Cr_2O_3$:Mg/ β-$Ga_2O_3$ p–n heterojunction diodes at 600 °C," Appl. Phys. Lett. **124**(15), 153504 (2024).

[33] B.J. Baliga, "Breakdown Voltage," in *Fundamentals of Power Semiconductor Devices*, edited by B.J. Baliga, (Springer International Publishing, Cham, 2019), pp. 89–170.

[34] X. Gonze, and C. Lee, "Dynamical matrices, Born effective charges, dielectric permittivity tensors, and interatomic force constants from density-functional perturbation theory," Phys. Rev. B **55**(16), 10355–10368 (1997).

[35] P.E. Blöchl, "Projector augmented-wave method," Phys. Rev. B **50**(24), 17953–17979 (1994).

[36] G. Kresse, and D. Joubert, "From ultrasoft pseudopotentials to the projector augmented-wave method," Phys. Rev. B **59**(3), 1758–1775 (1999).

[37] G. Kresse, and J. Furthmüller, "Efficient iterative schemes for *ab initio* total-energy calculations using a plane-wave basis set," Phys. Rev. B **54**(16), 11169–11186 (1996).

[38] J.P. Perdew, K. Burke, and M. Ernzerhof, "Generalized Gradient Approximation Made Simple," Phys. Rev. Lett. **77**(18), 3865–3868 (1996).

[39] Y. Liu, S. Roy, C. Peterson, A. Bhattacharyya, and S. Krishnamoorthy, "Ultra-low reverse leakage $NiO_x$/β-$Ga_2O_3$ heterojunction diode achieving breakdown voltage> 3 kV with plasma etch field-termination," AIP Advances **15**(1), (2025).

[40] P.H. Fang, and W.S. Brower, "Dielectric Constant of $Cr_2O_3$ Crystals," Phys. Rev. **129**(4), 1561–1561 (1963).

[41] X. Lu, X. Zhou, H. Jiang, K.W. Ng, Z. Chen, Y. Pei, K.M. Lau, and G. Wang, "1-kV Sputtered p-NiO/n-$Ga_2O_3$ Heterojunction Diodes With an Ultra-Low Leakage Current Below 1∼μA/$cm^2$," IEEE Electron Device Letters **41**(3), 449–452 (2020).

# Supplementary Material

# $Cr_2O_3/\beta\text{-}Ga_2O_3$ Heterojunction Diodes with Orientation-Dependent Breakdown Electric Field up to 12.9 MV/cm

Yizheng Liu[1,a], Haochen Wang[1], Carl Peterson[1], James S. Speck[1], Chris Van De Walle[1], and Sriram Krishnamoorthy[1,a]

[1]Materials Department, University of California Santa Barbara, Santa Barbara CA 93106, USA

a) Author(s) to whom correspondence should be addressed. Electronic mail: yizhengliu@ucsb.edu, sriramkrishnamoorthy@ucsb.edu

The capacitance-voltage (C-V) characteristics of both $n^+$ $\beta\text{-}Ga_2O_3$ Schottky barrier diodes (SBDs) and $Cr_2O_3/\beta\text{-}Ga_2O_3$ heterojunction diodes (HJDs) were measured at an AC frequency of 1 MHz on various orientations of $\beta\text{-}Ga_2O_3$.

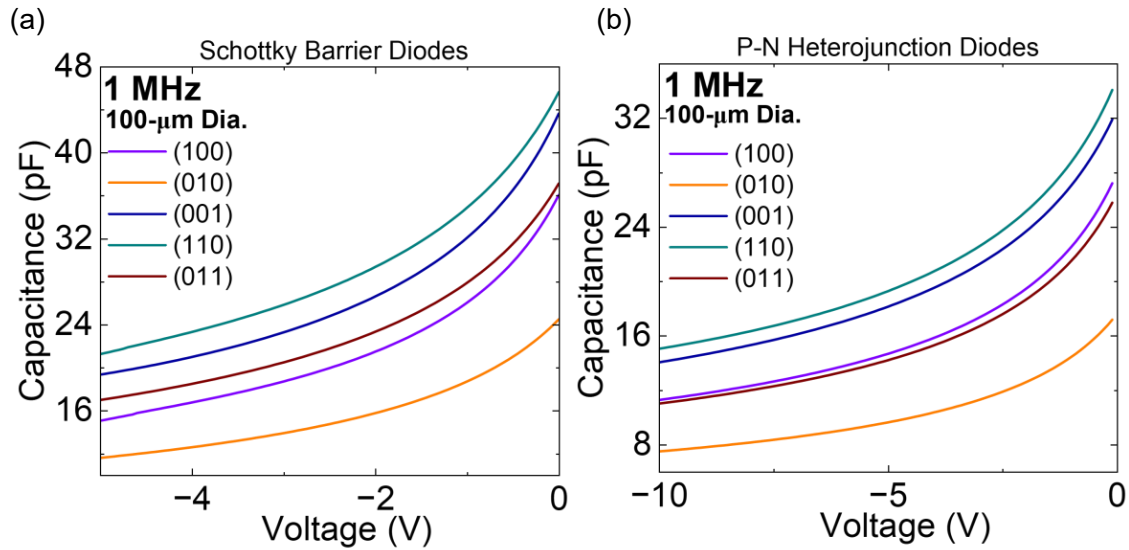

FIG.S1. (a) Capacitance vs. voltage characteristics of $n^+$ $\beta\text{-}Ga_2O_3$ Schottky barrier diodes at 1 MHz on 100-μm dia. devices in five orientations of bulk $\beta\text{-}Ga_2O_3$. (b) Capacitance vs. voltage characteristics of $Cr_2O_3/\beta\text{-}Ga_2O_3$ heterojunction diodes at 1 MHz on 100-μm dia. devices in five orientations of bulk $\beta\text{-}Ga_2O_3$.

The current density vs. voltage (J-V) characteristics of both n⁺ β-$Ga_2O_3$ Schottky barrier diodes (SBDs) and $Cr_2O_3$/β-$Ga_2O_3$ heterojunction diodes (HJDs) were measured on 100-μm dia. devices and were shown in linear and semi-log scale.

(a)

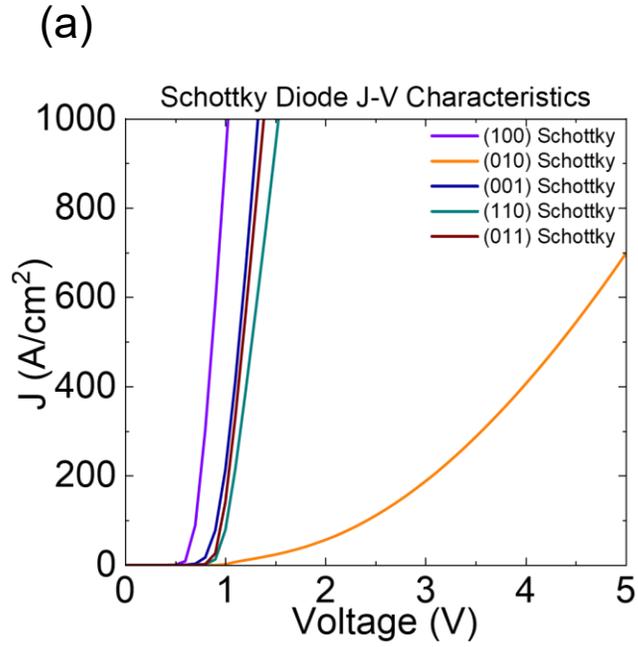

(b)

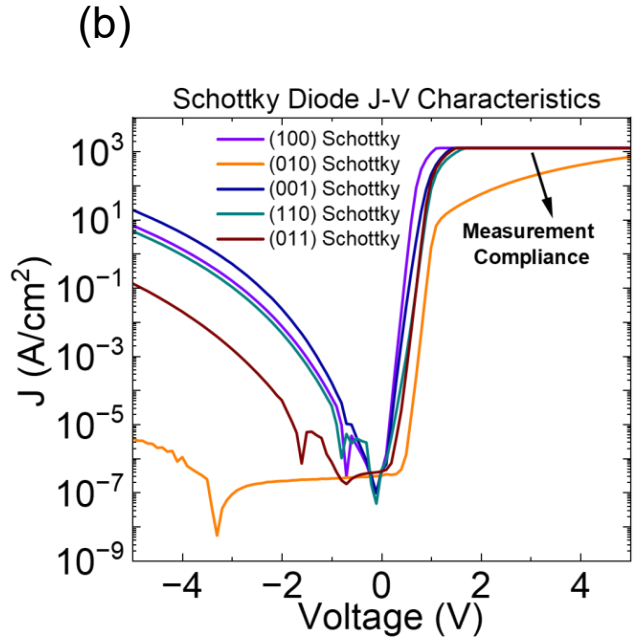

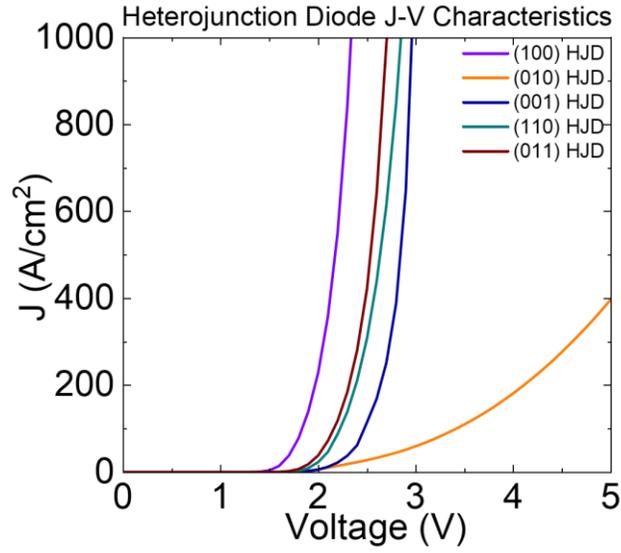

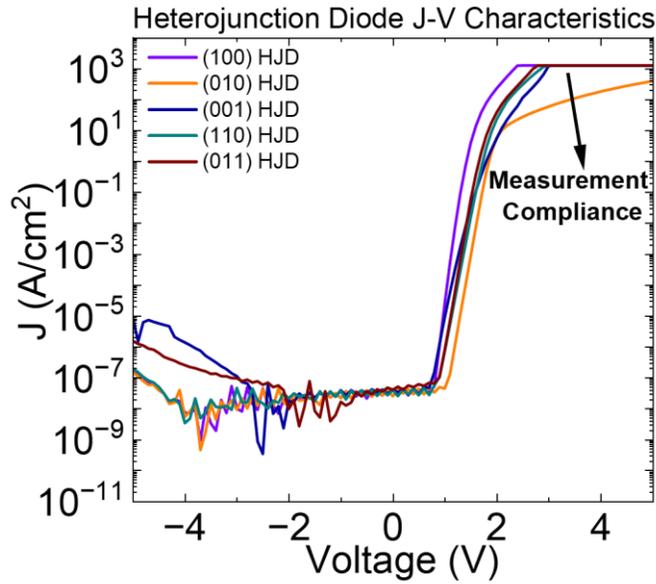

FIG.S2. (a) J-V characteristics of 100-μm dia. n$^+$ β-Ga$_2$O$_3$ vertical SBDs in five orientations of bulk β-Ga$_2$O$_3$. (b) J-V characteristics of 100-μm dia. n$^+$ β-Ga$_2$O$_3$ vertical SBDs in semi-log scale in five orientations of bulk β-Ga$_2$O$_3$. (c) J-V characteristics of 100-μm dia. Cr$_2$O$_3$/β-Ga$_2$O$_3$ vertical HJDs in five orientations of bulk β-Ga$_2$O$_3$. (d) J-V characteristics of 100-μm dia. n$^+$ β-Ga$_2$O$_3$ vertical HJDs in semi-log scale in five orientations of bulk β-Ga$_2$O$_3$.